# 5f Electron Induced Spin Transport by Sandwich-Type Phthalocyanine

Lu Xu[1], Ding Wang[1], Xiaobo Yuan[1], Dongfa Lan[1], Yu Zhu[2], Xiaobo Li[1], Weiyu Xie[1, *]

[1]*School of Physics and Optoelectronic Engineering, Hainan University, Haikou, 570228, China.*

[2]*College of Physics and Electronic Engineering, Hainan Normal University, Haikou 571158, China.*

Address correspondence to E-mail: wyxie@hainanu.edu.cn

**ABSTRACT:**

The rapid development of integrated circuits highlights the critical need for novel spintronic device designs. Sandwich-type phthalocyanine molecules, with their unique electrical and magnetic properties, show great potential in spintronic applications. Actinide elements, due to the strong interaction of their 5f electrons, can induce various exotic spin transport effects. In this study, we employed the non-equilibrium Green's function method combined with density functional theory (NEGF-DFT) to investigate the spin transport properties of the actinide sandwich phthalocyanine molecule $U(Pc)_2$. Electronic structure analysis indicates that the 5f electrons of uranium atom dominate its frontier orbital behavior. Transport property analyses reveal that when the bias voltage exceeds 0.4 V, the current increases significantly, due to an increase in the spin-up electron transmission peak, primarily contributed by 5f electrons. Our results underscore the dominant role of U-5f electrons in the spin transport of $U(Pc)_2$. This study aims to provide beneficial assistance for the development of actinide phthalocyanine molecular spintronic devices.

## Introduction

With the increasing demands for smaller device dimensions and improved performance in integrated circuits, many researchers view the design of new spintronic devices based on electronic spin properties as a crucial direction to overcome industrial bottle-necks [1-2], such as spin valves, spin filtering devices, and magnetic tunnel junctions (MTJs) [3-7]. In this context, the spin-dependent electronic structure and transport proper-ties of these devices can validate the feasibility of envisioned technologies for practical ap-plications. Thus, understanding the spin transport properties of these devices is not only of significant scientific importance but also addresses an increasingly urgent societal need for technological advancement.

Among the various materials capable of forming spintronic devices, phthalocyanine (Pc) molecules [8-12] stand out due to their large aromatic ring with 18 delocalized π-electrons and four nitrogen donors within a central cavity. These features make them efficient ligands, offering stability, metal specificity, ligand functionality, and reactivity [13]. Compounds derived from these molecules exhibit excellent transport properties and can function as negative differential resistors and spin filtering devices [14-22]. Through persistent research efforts, it has been found that Pcs can form metal complexes with most metallic elements [23]. Coordination with smaller atoms, such as Li or Fe, results in single-layer binuclear complexes or planar structural molecules, while coordination with larger heavy metal elements (e.g., Hg, actinides) leads to "half-sandwich" or "sandwich" type structures [6, 24]. Sandwich-type Pcs represent a significant subset, with numerous double-decker and triple-decker sandwich-type metal Pcs reported to date, particularly involving lanthanide and actinide elements [13, 25]. Due to their rich electronic structures, including spin polarization, these complexes are suitable for applications in spin field-effect transistors, giant magnetoresistance, high-density storage, and more [26-27].

For actinide elements, although the applications of actinide-related systems are restricted due to their toxicity and radioactivity, preliminary studies have already emphasized their great potential in electronic transport. Notably, studies of spin transport properties in actinide-encapsulated fullerene structures and carbon nanotubes have unveiled quantum phenomena such as spin filtering and spin switching, which are attributed to the presence of 5f electrons [28-29]. The exploration of sandwich-type Pc complexes with actinide elements has not progressed as rapidly as those with transition metals or lanthanides, it has garnered attention for several decades due to the unique interactions between actinide elements and Pc ligands [13]. In 1968, Lux et al. [30] first synthesized sandwich-type Pc complexes of the actinide elements Th(Pc)$_2$ and U(Pc)$_2$. In 1997, Kadish and colleagues proposed improved synthetic methods to increase the yield of Th(Pc)$_2$ and U(Pc)$_2$ [31]. By 1970, sandwich-type Pcs of Protactinium and Neptunium were also synthesized [32]. Subsequently, even super-uranium elements such as double-decker americium complexes Am(Pc)$_2$ were discovered by Moskalev et al. [33]. However, despite the early synthesis of actinide element sandwich-type Pcs, these compounds have not been extensively studied. Research into the spin transport properties of actinide element sandwich-type Pcs remains in its infancy, primarily due to the strong electron correlation and relativistic effects inherent to actinide elements, which result in complex valence electron behaviors and methodological challenges [34-36]. Our preliminary studies on actinide-encapsulated systems have demonstrated their potential in electron transport, such as the spin filtering effects observed in a series of actinide-encapsulated carbon nanotubes [28]. Recently, systematic investigations of An-encapsulated clusters have also shown that actinide encapsulation significantly enhances electronic transport capabilities [37].

In this study, we systematically investigated the electronic structure properties of the actinide sandwich-type Pc cluster U(Pc)$_2$. Additionally, employing non-equilibrium Green's function combined with density functional theory (NEGF-DFT), we explored the spin transport properties of U(Pc)$_2$ clusters, revealing a close relationship between spin transport properties and the 5f electrons of uranium. These results deepen our under-standing of the role of actinide elements in promoting spin transport and contribute to the development of spintronic molecular devices based on sandwich-type Pcs.

## Computational Models and Methods

U(Pc)$_2$ was thoroughly optimized using the B3LYP functional [38], which has demonstrated effectiveness in phthalocyanine molecular calculations [39-40]. Additionally, the geometric configurations and energies of U(Pc)$_2$ were computed using the M06-2X [41], PBE0 [42], and PBE [43] functionals. To account for relativistic effects, uranium was treated with a small-core relativistic effective core potential (RECP) comprising 60 core electrons, and a valence electron basis set of (14s13p10d8f6g)/[10s9p5d4f3g] [44]. For other elements, computational methods employed the double-$\zeta$ basis set 6-31G* with d polarization functions and the triple-$\zeta$ valence polarization basis set (def-TZVP). Vibrational frequency calculations were performed for all optimized structures at the same level to ensure the absence of imaginary frequencies. Energy decomposition analysis (EDA) was conducted using the sobEDA method, integrated within the Multiwfn 3.6 program [45-46], which has been shown to be effective for actinide compounds [47], to further elucidate the intrinsic nature of orbital interactions. The fragment interaction energy ($\Delta E_{int}$) was decomposed into four components:

$$\Delta E_{int} = \Delta E_{elstat} + \Delta E_c + \Delta E_{xrep} + \Delta E_{orb}$$

where $\triangle E_{xrep}$ denotes exchange-correlation contributions, $\triangle E_c$ represents Coulombic effects, $\triangle E_{elstat}$ corresponds to classical electrostatic interactions, and $\triangle E_{orb}$ accounts for orbital interactions. All computations were performed using the Gaussian package [48].

To calculate the spin transport properties, the non-equilibrium Green's function combined with density functional theory (NEGF-DFT) method was used to analyze the electronic transport characteristics of the system [49]. Utilizing the optimized stable structure of U(Pc)$_2$, Au electrodes were applied to construct a molecular spintronics device model, as these electrode materials are widely used in transport property calculations for such molecules [14, 15, 50-51]. Since there were no direct chemical bonds between the U(Pc)$_2$ molecules and the gold electrodes, slight modifications were made to the molecular structure by replacing hydrogen atoms at the connection points with sulfur atoms, enabling U(Pc)$_2$ to attach to the electrodes through these sulfur atoms. In the calculations, the PBE functional was employed, utilizing double-$\zeta$ polarized atomic orbital basis sets (DZP) for Au, C, N, S, and H atoms, while pseudopotentials developed in previous studies were used to describe the core electron behavior of U atoms. For self-consistency and transmission coefficient calculations, a k-point grid of 1×1×1 was employed. Initially, self-consistent potentials were computed for both electrode regions, followed by self-consistent calculations for the central scattering region of the device. It should be noted that NEGF theory assumes no interface scattering in the middle region, i.e., between the electrode and the central region. To ensure a smooth transition of device potentials with the electrode portions, sufficient layers of electrode material were retained as buffer layers. For accuracy, a vacuum length of 20 Å was maintained, extending 4 layers beyond the electrode as buffer layers. Transport calculations were performed using the Nanodcal code [52]. The schematic structure of the model is shown in Figure 3. The spin-correlated current of this system was obtained using the Landauer-Büttiker formula [53]:

$$I(V) = \frac{2e^2}{h} \int_{\mu_L}^{\mu_R} \left( T(E) [f_L(E) - f_R(E)] \right) dE$$

where $\mu_L$ and $\mu_R$ is the chemical potential of left and right electrodes, respectively, while $f_L(E)$ and $f_R(E)$ are electronic Fermi-Dirac distribution functions of left/right electrodes [54]. $T(E)$ is the transmission probability which can be calculated as:

$$T(E) = Tr\left[\Gamma_R(E,V)G^R(E,V)\Gamma_L(E,V)G^A(E,V)\right]$$

where $G^R$ and $G^A$ are the retarded and advanced Green's function, respectively, while $\Gamma_L$ and $\Gamma_R$ is the imaginary part of the self-energies known as coupling functions induced by interaction of channel area with electrodes.

## Results and Discussion

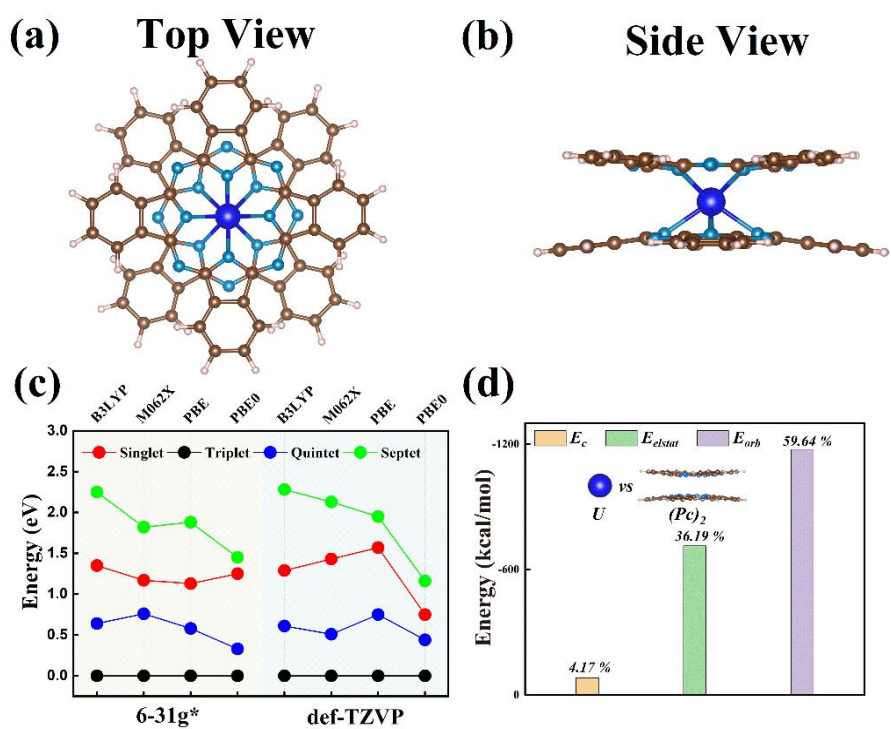

**Figure 1.** (a) and (b) Geometric configurations of different views of U(Pc)$_2$. The spheres represent different elements: brown for carbon (C), sky blue for nitrogen (N), dark blue for uranium (U), and pink for hydrogen (H). Coordinates of all optimized structures are provided in Supplementary Material Tables S6). (c) The relative energies of U(Pc)$_2$ between different electronic states performed by B3LYP, M062X, PBE and PBE0 functional with different basis sets (It should be noted that different basis sets were targeted at elements other than the U atom. U atom was always used the RECP pseudopotential basis set). (d) EDA results for U(Pc)$_2$ at the B3LYP/def-TZVP/RECP level.

After extensive structural relaxation, the stable configuration of U(Pc)$_2$ was obtained. Due to the large atomic radius of uranium, its combination with Pc molecules exclusively forms a "sandwich" structure. In this configuration, the uranium atom is positioned between two Pc molecules, each of which contains four N$_4$ sites capable of bonding with the uranium atom, resulting in a bilayer sandwich-type Pc molecule (Fig. 1(a), (b)). It was observed that the originally planar Pc molecules undergo bending during the binding process, ultimately adopting a "butterfly" shape. This deformation occurs because the uranium atom is drawn towards the large rings of the two Pc molecules, causing the edges of the U(Pc)$_2$ structure to bend in opposite directions. This destroys the symmetry of the original single-layer Pc molecule. Therefore, in this work, the structure does not consider symmetry. Computational analysis reveals an average U-N bond length of 2.43 Å, indicating a strong interaction between the uranium atom and the Pc molecules.

Regarding the electronic structure of U(Pc)$_2$, the calculations indicate that the ground state is a triplet (Figure 1(c) and Supplementary Material Tables S1-S4). Further analysis reveals that, across different combinations of functionals and basis sets, the energy order of U(Pc)$_2$ electronic states consistently follows the pattern: triplet < quintet < singlet < septet. The energy difference between the ground state and the quintet state is minimal. However, there are significant energy gaps between the ground state and both the singlet and septet states, suggesting substantial electronic structural changes in these two states, which may indicate improper electron occupation. Due to the presence of actinide elements, the consideration of spin-orbit coupling cannot be ignored. Therefore, for verification purposes, we performed single-point energy calculations of SO-DFT at the PBE/def-TZVP level (Supplementary Material Tables S5). The calculation results indicate that when the spin-orbit coupling effect is taken into account, the ground state remains a triplet. However, the energy order of each electronic state has changed. The closed-shell singlet becomes the state with the highest energy, and the energy differences among the three high-spin states become very small. The quintet and septet states are only 0.05 eV and 0.07 eV higher in relative energy than the triplet state. Therefore, we will only analyze the structure of the most stable triplet state below.

To further investigate the bonding characteristics between the uranium atom and the Pc molecules in the U(Pc)$_2$ structure, energy decomposition analysis (EDA) was performed. The uranium atom was designated as Fragment 1, and the two Pc molecules as Fragment 2. This method allows for a quantitative breakdown of the interactions between uranium and the Pc molecules into physically meaningful components. As shown in Figure 1(d), the attractive components were further subdivided into electrostatic attraction ($\Delta E_{elstat}$, -713.52 kcal/mol), orbital interaction ($\Delta E_{orb}$, -1175.43 kcal/mol), and Coulomb correction ($\Delta E_c$, -82.34 kcal/mol). The denotes exchange-correlation contributions ($\triangle$

E$_{xrep}$, 1481.31 kcal/mol). The interaction between uranium and the Pc molecules is primarily governed by orbital interactions and electrostatic attractions, with the orbital interactions likely originating from the eight U-N bonds. These findings clarify how the presence of the uranium atom contributes to the strong binding characteristic of the sandwich Pc structure.

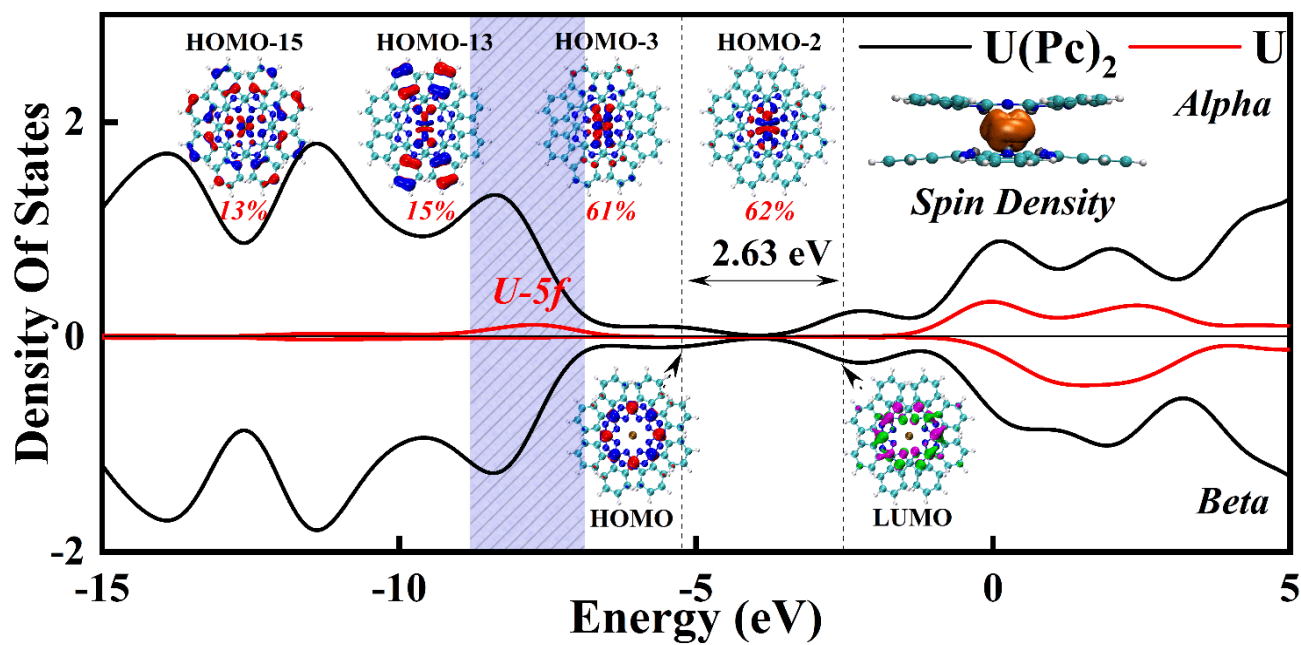

**Figure 2.** Density of states diagram of U(Pc)$_2$ at the M062x/def-TZVP/RECP level. DOS diagram uses a Gaussian function to expand the discrete orbitals into the form of peaks, and the full width at half maximum (FWHM) is set to 1.36 eV. The dashed lines show the positions of the HOMO and LUMO energy levels. The shaded blue region represents the primary contribution area of U-5f electronic component. The red percentages indicate the contribution of U-5f electrons in the molecular orbitals. The isosurface value for the molecular orbital diagrams is set at 0.02, while the isosurface for the spin density diagram is set at 0.005.

Understanding the frontier orbitals of U(Pc)$_2$ is crucial for exploring its electronic transport properties, as interactions between these orbitals predominantly govern electron transfer processes. Figure 2 presents the density of states (DOS) of U(Pc)$_2$. Due to the triplet state of U(Pc)$_2$, the DOS exhibits asymmetric peaks for spin-up (α) and spin-down (β) electrons. The calculated HOMO-LUMO band gap of U(Pc)$_2$ is 2.63 eV, which is within the typical range for semiconductor band gaps. The spin density diagram and net spin electron distribution of U(Pc)$_2$ shows that the two net spin electrons are predominantly localized on the uranium atom, and they are fully contributed by the U-5f shell, suggesting a strong correlation between the spin transport properties and the uranium atom in U(Pc)$_2$.

Orbital composition calculations adopt the Hirshfeld method, and the results indicate that the HOMO is entirely derived from the p orbitals of carbon atoms connected to the four nitrogen vacancies in the two Pc molecules. The LUMO is composed of approximately 70% p orbitals from carbon atoms, with the remaining contributions from nitrogen p orbitals. HOMO-1 exhibits a composition similar to that of the HOMO. For HOMO-2 and HOMO-3, the uranium atom's 5f electrons contribute significantly, with contributions of 62% and 61%, respectively. Notably, this behavior aligns with previous studies on U@$C_{28}$, where two singly occupied orbitals (HOMO and HOMO-1) are predominantly contributed by carbon atoms, while lower-energy orbitals begin to involve uranium electron participation [55-56]. Figure 2 highlights the energy regions where U-5f electrons contribute to molecular orbital formation, focusing on several representative molecular orbitals such as HOMO-2, HOMO-3, HOMO-13 and HOMO-15. These orbitals predominantly involve U-5f electron contributions, though contributions from other orbitals are not excluded. The shapes of these orbitals are characteristic of f orbital configurations. Additionally, in the unoccupied orbitals above the LUMO, the uranium atom also makes a significant contribution, positively influencing spin transport properties. Regarding the U-6d electrons, we found that they are delocalized on various molecular orbitals and are not localized on a certain orbital. In summary, DOS analysis reveals that the uranium atom dominates the behavior of its frontier molecular orbitals, contributing to the unique spin transport properties induced by actinide elements in U(Pc)$_2$.

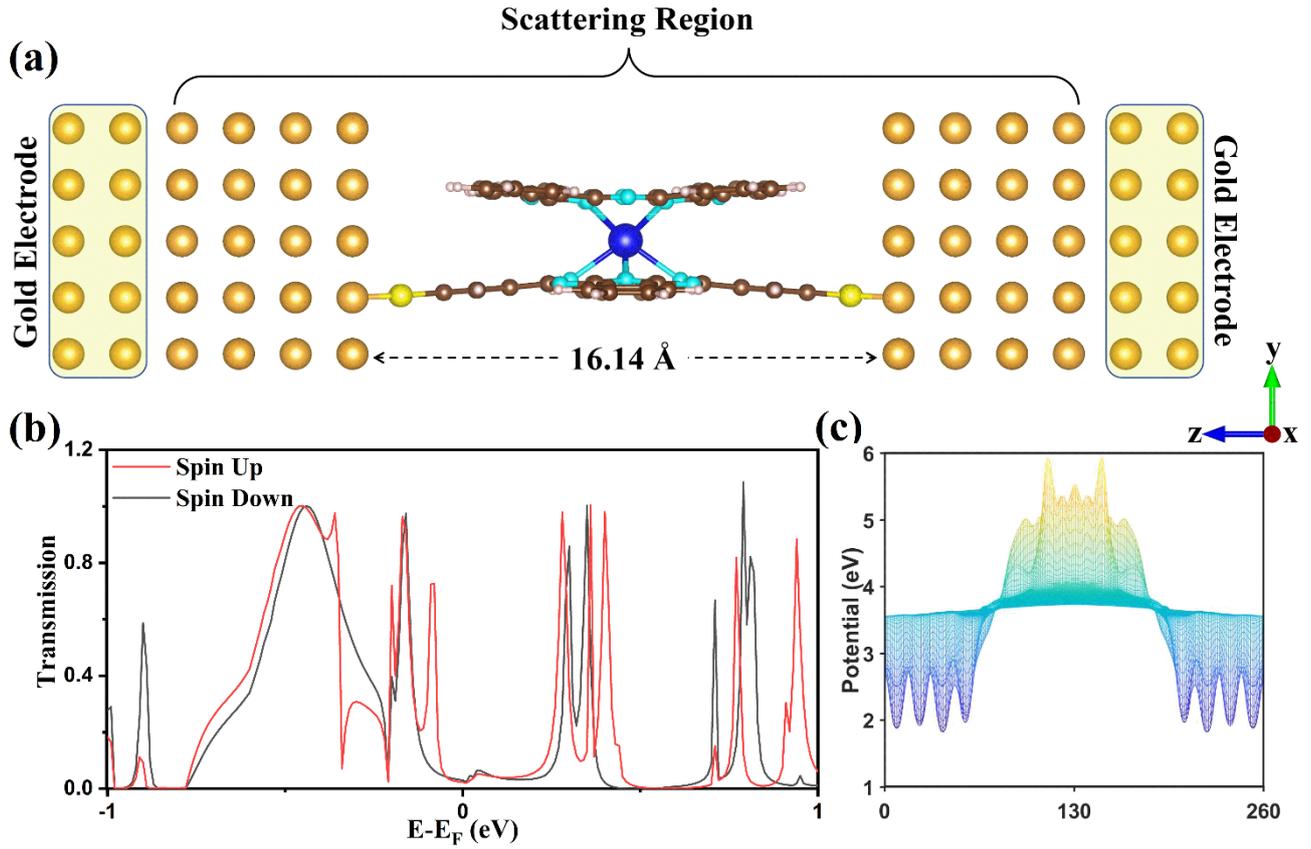

**Figure 3.** (a) Transport model: U(Pc)$_2$-LR. The transport direction is along the Z-axis. The brown, sky blue, dark blue, pink, and yellow spheres represent the C, N, U, H, and S elements, respectively. (b) Zero-bias transmission spectra of U(Pc)$_2$-LR systems. For clarity, the Fermi level has been adjusted to $E_F = 0$. (c) Potential distribution function of U(Pc)$_2$-LR, with the horizontal axis representing the transport direction.

To investigate the spin transport properties of U(Pc)$_2$, the NEGF-DFT method was employed. As shown in Figure 3(a), a left-right electrode transport model with gold contacts, U(Pc)$_2$-LR, was established. To validate the efficacy of the device model's shielding effects, the real-space distribution of the system's potential energy along the transport direction was computed, as shown in Figure 3(c). The potential energy distribution near the buffer region close to the electrodes remains unchanged, indicating effective system shielding.

The transmission coefficient T(E) is a critical physical parameter for evaluating single-molecule devices. Initially, the equilibrium electron transmission function for U(Pc)$_2$-LR was computed. As depicted in Figure 3(b), notable differences are observed in the spin-up and spin-down transmission spectra within the energy range of -1 to 1 eV, with no prominent peaks near the Fermi level. Between -0.25 and 0 eV, a peak is evident in the spin-up electron projection, whereas spin-down electrons do not exhibit such a feature. At lower energies (-1 to -0.25 eV), there is negligible disparity in projection coefficients between spin-up and spin-down electrons. In the range of 0 to 0.5 eV, robust projection peaks are observed for both spin-up and spin-down electrons, with spin-up dominance. Due to the focus on the transmission function near the Fermi level in equilibrium states, U(Pc)$_2$ is anticipated to exhibit weak conductivity. The equilibrium conductance, calculated using the transmission coefficient T(E$_F$) at the Fermi level, is 0.05 G$_0$. In comparison with the equilibrium conductance of a similar phthalocyanine transport system (CuPc: approximately 0.006 G$_0$), the conductance of U(Pc)$_2$ has been greatly enhanced [57]. This analysis suggests that because of the finite band gap of U(Pc)$_2$, there are no molecular orbitals precisely at the Fermi level. However, the presence of strong projection peaks on either side of the Fermi level suggests potentially superior electronic spin transport properties in non-equilibrium states for this device.

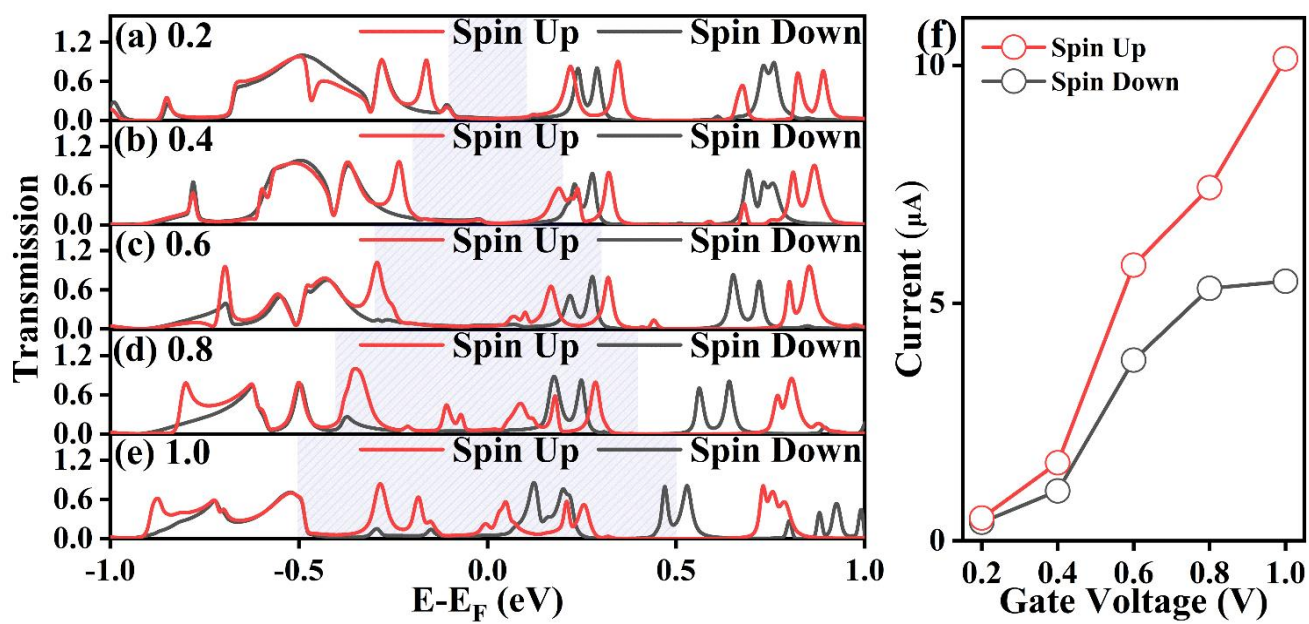

**Figure 4.** (a) to (e) Transmission spectra of U(Pc)$_2$ systems under 0.2-1.0 V bias. The shadows are the integral area of the bias voltage energy window. (f) I-V curve of U(Pc)$_2$.

Analysis of the density of states (DOS) for U(Pc)$_2$ reveals significant contributions from U-5f electrons around both sides of the Fermi level. To explore the spin transport properties of U(Pc)$_2$ under applied biases ranging from 0.2 to 1.0 V, we plotted the corresponding transmission functions. Specifically, voltages from 0.1 to 0.5 V were applied to one electrode, while corresponding voltages from -0.1 to -0.5 V were applied to the other, resulting in an overall bias range of 0.2 to 1.0 V. It is important to note that the system's energy levels shift in response to the applied bias voltage, modulating its transport characteristics. Figure 4(a) to 4(e) illustrate the transmission spectra under these biases, with shaded regions highlighting the integrated energy windows for each bias to facilitate direct comparison.

Overall, the transmission peak shifts to the left with increasing bias. At 0.2 and 0.4 V biases, although the bias window expands, only a few transmission peaks are present, with a larger integrated area for spin-up electrons. Starting from a 0.6 V bias, energy level shifts and widened bias windows make transmission peaks for spin-up electrons on the left and both spin-up and spin-down peaks on the right of the Fermi level more prominent. Particularly noteworthy are the increased transmission peaks for spin-up electrons on the left at 0.6 and 0.8 V biases, which dominate the electron transport behavior. At 1.0 V bias, an increase in the number of spin-up transmission peaks is observed, with a spin-up electron transmission peak appearing exactly at the Fermi level due to energy level tuning, highlighting differences in the integrated regions between spin-up and spin-down electrons.

To verify these results, we analyze the current-voltage (I-V) curves, as shown in Figure 4(f). We find that the spin current exhibits a nonlinear relationship with the applied bias voltage (V), with the overall current (I) increasing as V increases. At lower biases (below 0.4 V), the spin-up and spin-down current components are both minimal and nearly identical. As the bias is increased further (above 0.6 V), the current values rise sharply. This disparity widens with increasing bias, consistent with the transmission spectra results. Thus, spin-up electrons play a crucial role in the non-equilibrium spin transport of U(Pc)$_2$, which can be attributed to the influence of U-5f electron orbitals enhancing the spin-up current component. Compared with the I-V curve of a similar system such as the Au-FePc-Au transport model, the current value of U(Pc)$_2$ has been increased by about an order of magnitude [58]. In comparison with the studies of a series of MPc (M = Ni, Fe, Co, Mn, Cr) connected to gold electrodes [59], we find that the current value of U(Pc)$_2$ at the same bias voltage is also much larger. All these reflect the advantages of actinide phthalocyanines in spin transport.

## Conclusion

In this study, by using the NEGF-DFT method, we conducted a comprehensive exploration into the geometry, electronic structure, and spin transport characteristics of U(Pc)$_2$. The ground state of U(Pc)$_2$ is triplet, and intriguingly, all of the net spin electrons are furnished by the U-5f electrons. Through an in-depth density of states (DOS) analysis, it became evident that the 5f electrons of the uranium atom wield preponderant influence over its frontier molecular orbital behavior. Building on these foundational electronic structure insights, we delved further into the spin transport properties of U(Pc)$_2$. A meticulous analysis of the I-V curves and transmission spectra disclosed that once the bias voltage surpasses 0.4 V, a marked augmentation in current is witnessed, with spin-up electrons being the principal contributors. Moreover, as the bias voltage escalates, the disparity between the spin-up and spin-down electron populations becomes increasingly conspicuous. These revelations unequivocally underscore the pivotal role played by the 5f electrons in dictating the spin transport properties of U(Pc)$_2$. Admittedly, our current exploration of the spin transport properties of actinide phthalocyanines is merely in its nascent stage. Nevertheless, we ardently anticipate that future research efforts will focus more intensively on the related systems of actinide elements. Such investigations hold the promise of not only facilitating the fine-tuning and inventive design of molecular nano-devices but also catalyzing a more profound comprehension of the enigmatic electronic properties of 5f elements.

*Data Availability*: The data that support the findings of this study are available from the authors upon request.

*Competing Interests*: The authors declare that there are no competing interests.

*Acknowledgments*: This work was supported by the National Natural Science Foundation of China (No. 12204134), Hainan Provincial Natural Science Foundation of China (123RC397, 121QN167 and 124QN247) and the Start-up Research Foundation of Hainan University (KYQD(ZR)-20093). We thank HZWTECH company for help and discussions regarding this study.

## References

[1] Li, D.; Yu, G. Innovation of Materials, Devices, And Functionalized Interfaces in Organic

Spintronics. *Adv. Funct. Mater.* **2021**, *31*, 2100550.

[2] Guo, Y. Q.; Zhang, X.; Huang, Z.; Chen, J. Y.; Luo, Z. J.; Zhang, J.; Li, J. F.; Zhang, Z. W.; Zhao, J. K.; Han X. F.; Wu, H. Quantum Materials for Spintronic Applications. *npj Spintronics* **2024**, *2*, 36.

[3] Xiong, Z. H.; Wu, D.; Vardeny, Z. V.; Shi, J. Giant Magnetoresistance in Organic Spin-Valves. *Nature* **2004**, *427*, 821-824.

[4] Su, T. A.; Neupane, M.; Steigerwald, M. L.; Venkataraman, L.; Nuckolls C. Chemical Principles of Single-Molecule Electronics. *Nat. Rev. Mater.* **2016**, *1*, 16002.

[5] Gehring, P.; Thijssen, J. M.; van der Zant, H. S. J. Single-Molecule Quantum-Transport Phenomena in Break Junctions. *Nat Rev Phys.* **2019**, *1*, 381-396.

[6] Doud, E. A.; Voevodin, A.; Hochuli, T. J.; Champsaur, A. M.; Nuckolls, C.; Roy X. Superatoms in Materials Science. *Nat. Rev. Mater.* **2020**, *5*, 371-387.

[7] Dong, X.; Jia, X.; Yan, Z.; Shen, X.; Li, Z.; Qiao, Z.; Xu, X. Spin Transport Properties of $MnBi_2Te_4$-Based Magnetic Tunnel Junctions. *Chin. Phys. Lett.* **2023**, *40*, 087301.

[8] Dolphin, D.; Felton, R. H. Biochemical Significance of Porphyrin. pi. Cation Radicals. *Acc. Chem. Res.* **1974**, *7*, 26-32.

[9] Uyeda, N.; Kobayashi, T.; Suito, E.; Harada, Y.; Watanabe M. Molecular Image Resolution in Electron Microscopy. *J. Appl. Phys.* **1972**, *43*, 5181-5189.

[10] Li, M.; Han, B.; Li, S.; Zhang, Q.; Zhang, E.; Gong, L.; Jiang, J. Constructing 2D Phthalocyanine Covalent Organic Framework with Enhanced Stability and Conductivity via Interlayer Hydrogen Bonding as Electrocatalyst for $CO_2$ Reduction. *Small* **2024**, 2310147.

[11] Kumar, A.; Vashistha, V. K.; Das D. K. Recent Development on Metal Phthalocyanines Based Materials for Energy Conversion and Storage Applications. *Coordin. Chem. Rev.* **2021**, *431*, 213678.

[12] Barraud, C.; Bouzehouane, K.; Deranlot, C.; Kim, D. J.; Rakshit, R.; Shi, S.; Arabski, J.; Bowen, M.; Beaurepaire, E.; Boukari, S.; Petroff, F.; Seneor, P.; Mattana, R. Phthalocyanine Based Molecular Spintronic Devices. *Dalton Trans.* **2016**, *45*, 16694-16699.

[13] VargasZúiga, G. I.; Boreen, M. A.; Mangel, D. N.; Arnold, J.; Sessler, J. L. Porphyrinoid Actinide Complexes. *Chem. Soc. Rev.* **2022**, *51*, 3735-3758.

[14] Douand, K. P.; Kaun, C. C. Conductance Switching of a Phthalocyanine Molecule on An Insulating Surface. *Front. Phys.* **2017**, *12*, 127303.

[15] Vincent R, Klyatskaya S, Ruben M, et al. Electronic read-out of a single nuclear spin using a molecular spin transistor. *Nature*, **2012**, *488*, 357-360.

[16] Liu, X. J.; Wang, T. J.; Niu, L.; Wang, Y.; Zhang, Q.; Yin, H. T.; Magnetic Proximity Magnetoresistance and Spin Filtering Effect in A Binuclear Ferric Phthalocyanine from First Principles. *J. Phys. D Appl. Phys.* **2020**, *53*, 035305.

[17] Chen, Y. L.; Qu, S. Y.; Song, Q. F.; Shi, W.; Li, H.; Yao, Q.; Chen, L. D. Synergistically Optimized Electrical and Thermal Transport Properties in Copper Phthalocyanine-Based Organic Small Molecule with Nanoscale Phase Separations. *ACS Appl. Mater. Interfaces* **2021**, *13*, 15064-15072.

[18] Godfrin C, Ferhat A, Ballou R, et al. Operating quantum states in single magnetic molecules: implementation of Grover's quantum algorithm. *Phys. Rev. Lett.* **2017**, *119*, 187702.

[19] Huang, J.; Xu, K.; Lei, S.; Su, H.; Yang, S.; Li Q.; Yang, J. Iron-Phthalocyanine Molecular Junction with High Spin Filter Efficiency and Negative Differential Resistance. *J. Chem. Phys.* **2012**, *136*, 064707.

[20] Li, J.; Hu, J.; Wang, H.; Wu, R. Q. Rhenium-Phthalocyanine Molecular Nanojunction with High Magnetic Anisotropy and High Spin Filtering Efficiency. *Appl. Phys. Lett.* **2015**, *107*, 032404.

[21] Zhao, W.; Zou, D.; Yang, C. L.; Sun, Z. Spin Transport Properties and Spin Logic Gates in Manganese Phthalocyanine-Based Molecular Combinational Circuits. *J. Mater. Chem. C* **2017**, *5*, 8862-8868.

[22] Tao, L. L.; Wang, J. Giant Magnetoresistance and Perfect Spin Filter Effects in Manganese Phthalocyanine Based Molecular Junctions. *Nanoscale* **2017**, *9*, 12684-12689.

[23] McKEOWN, N B. The Synthesis of symmetrical 98 phthalocyanines. *The Porphyrin Handbook: Phthalocyanines: Synthesis*, **2000**, *15*, 61-369.

[24] Inabe, T.; Tajima, H. Phthalocyanines versatile components of molecular conductors. *Chem. Rev.* **2004**, *104*, 5503-5534.

[25] Brewster, J. T.; Zafar, H.; Root, H. D.; Thiabaud, G. D.; Sessler, J. L. Porphyrinoid f-Element Complexes. *Inorg. Chem.* **2019**, *59*, 32-47.

[26] Gehring, P.; Thijssen, J. M.; van der Zant, H. S. J. Single-Molecule Quantum-Transport Phenomena in Break Junctions. *Nat. Rev. Phys.* **2019**, *1*, 381-396.

[27] Tabata, C.; Watanabe, H.; Shirasaki, K.; Sunaga, A.; Fukuda, T.; Li, D.; Yamamura, T. Crystallographic and/or Magnetic Properties of Neutral and Cationic Uranium (IV) Sandwiched Phthalocyanine Complexes. *J. Mol. Struct.* **2023**, *1277*, 134870.

[28] Wang, J.; Gong, K.; Lu, F.; Xie, W.; Zhu, Y.; Wang, Z. Electronic Transport Inhibiting of Carbon Nanotubes by 5f Elements. Adv. Theory Simul. 2020, 3, 1900226.

[29] Xu K., Huang J., Lei S., et al. Efficient organometallic spin filter based on Europium-cyclooctatetraene wire. J. Chem. Phys. 2009 131, 104704.

[30] Lux, F.; Dempf, D.; Graw, D. Diphthalocyaninoto--Thorium (IV) and Uranium (IV). *Angew. Chem. Int. Ed. Engl.* **1968**, *7*, 819-820.

[31] Ng, D. K. P.; Jiang, J. Sandwich-Type Heteroleptic Phthalocyaninato and Porphyrinato Metal Complexes. *Chem. Soc. Rev.* **1997**, *26*, 433-442.

[32] Lux, F.; Ammentorp-Schmidt, F.; Dempf, D.; Graw, D.; Hagenberg, W. Radiochemical Investigations on Phthalocyaninto Actinide Complexes. Detection of the Bis (Phthalocyaninato) Complexes of Protactinium and Neptunium. *Radiochim. Acta.* **1970**, *14*, 57-61.

[33] Moskalev, P. N.; Shapkin, G. N.; Darovskih, A. N. Synthesis and Properties of Electrochemically Oxidized Diphthalocyanines of Rare Earths and Americium. *Z. Neorg. Khim.* **1979**, *24*, 340-346.

[34] Göetzke, L.; Schaper, G.; März, J.; Kaden, P.; Huittinen, N.; Stumpf, T.; Kammerlander, K. K. K.; Brunner, E.; Hahn, P.; Mehnert, A.; Kersting, B.; Henle, T.; Lindoy, L. F.; Zanoni, G.; Weigand J. J. Coordination Chemistry of F-Block Metal Ions with Ligands Bearing Bio-Relevant Functional Groups. *Coordin. Chem. Rev.* **2019**, *386*, 267-309.

[35] Marks, T. J.; Kolb, J. R. Covalent Transition Metal, Lanthanide, and Actinide Tetrahydroborate Complexes. *Chem. Rev.* **1977**, *77*, 263-293.

[36] Wang, D.; van Gunsteren, W. F.; Chai, Z. Recent Advances in Computational Actinoid Chemistry. *Chem. Soc. Rev.* **2012**, *41*, 5836.

[37] Gao, Y.; Xie, W.; Wang, B.; Schreckenbach, G.; Govorov, A. O.; Li, X.; Wang, Z. M. Observing the Role of Electron Delocalization in Electronic Transport by Incorporating Actinides into Ligated Metal-Chalcogenide Superatoms. *Langmuir* **2024**, *40*, 15023-15030.

[38] Lee, C.; Yang, W.; Parr, R. G. Development of The Colle-Salvetti Correlation-Energy Formula into A Functional of The Electron-Density. Phys. Rev. B 1988, 37, 785-789.

[39] Kobayashi, N.; Fukuda, T. Chapter 1 - Recent Progress in Phthalocyanine Chemistry: Synthesis and Characterization. Func-tional dyes 2006, 1-45.

[40] Visser, S. P. D.; Stillman, M. J. Challenging Density Functional Theory Calculations with Hemes and Porphyrins. Int. J. Mol. Sci. 2016, 17, 519.

[41] Zhao, Y.; Truhlar, D. G. The M06 Suite of Density Functionals for Main Group Thermochemistry, Thermochemical Kinetics, Noncovalent Interactions, Excited States, And Transition Elements: Two New Functionals and Systematic Testing of Four M06-Class Functionals And 12 Other Functionals. Theor. Chem. Acc. 2008, 120, 215-241.

[42] Adamo, C.; Barone, V. Toward Reliable Density Functional Methods Without Adjustable Parameters: The PBE0 Model. J. Chem. Phys. 1999, 110, 6158-6170.

[43] Perdew, J. P.; Burke, K.; Ernzerhof, M. Generalized Gradient Approximation Made Simple. Phys. Rev. Lett. 1996, 77, 3865-3868.

[44] Cao, X.; Dolg, M. Segmented Contraction Scheme for Small-Core Actinide Pseudopotential Basis Sets. J. Mol. Struc.: THEOCHEM. 2004, 673, 203-209.

[45] Lu T, Chen F. Multiwfn: A multifunctional wavefunction analyzer. J. Comput. Chem. 2012, 33, 580-592.

[46] Lu T. A comprehensive electron wavefunction analysis toolbox for chemists, Multiwfn. J. Chem. Phys. 2024, 161(8).

[47] Du, J.; Jiang, G.; Chen, D. High Coordination Numbers of Actinides (An) in AnC13+ Rings (An=Th and U). Inorg. Chem. 2023, 62, 20488-20495.

[48] M. J. Frisch, G. W. Trucks, H. B. Schlegel, G. E. Scuseria, M. A. Robb, J. R. Cheeseman, G. Scalmani, V. Barone, B. Mennucci, G. A. Petersson, H. Nakatsuji, M. Caricato, X. Li, H. P. Hratchian, A. F. Izmaylov, J. Bloino, G. Zheng, J. L. Sonnenberg, M. Hada, M. Ehara, K. Toyota, R. Fukuda, J. Hasegawa, M. Ishida, T. Nakajima, Y. Honda, O. Kitao, H. Nakai, T. Vreven, J. A. Montgomery, Jr., J. E. Peralta, F. Ogliaro, M. Bearpark, J. J. Heyd, E. Brothers, K. N. Kudin, V. N. Staroverov, R. Kobayashi, J. Normand, K. Raghavachari, A. Rendell, J. C. Burant, S. S. Iyengar, J. Tomasi, M. Cossi, N. Rega, J. M. Millam, M. Klene, J. E. Knox, J. B. Cross, V. Bakken, C. Adamo, J. Jaramillo, R. Gomperts, R. E. Stratmann, O. Yazyev, A. J. Austin, R. Cammi, C. Pomelli, J. W. Ochterski, R. L. Martin, K. Morokuma, V. G. Zakrzewski, G. A. Voth, P. Salvador, J. J. Dannenberg, S. Dapprich, A. D. Daniels, Ö. Farkas, J. B. Foresman, J. V. Ortiz, J. Cioslowski, D. J. Fox 2009 Inc. Wallingford CT 201.

[49] Xue, Y.; Datta, S.; Ratner, M. A. First-Principles Based Matrix Green's Function Approach to Molecular Electronic Devices: General Formalism. Chem. Phys. 2002, 281, 151-170.

[50] Song, Y.; Wang, C. K.; Chen, G.; Zhang, G. P. A First-Principles Study of Phthalocyanine-Based Multifunctional Spintronic Molecular Devices. Phys. Chem. Chem. Phys. 2021, 23, 18760-18769.

[51] Liu, X.; Wang, T.; Niu, L.; Wang, Y.; Zhang, Q.; Yin, H. Magnetic Proximity, Magnetoresistance and Spin Filtering Effect in A Binuclear Ferric Phthalocyanine from First Principles. J. Phys. D Appl. Phys. 2019, 53, 035305.

[52] Jeremy, T.; Guo, H.; Wang, J. Ab Initio Modeling of Quantum Transport Properties of Molecular Electronic Devices. Phys. Rev. B 2001, 63, 245407.

[53] Datta, S. Exclusion Principle and The Landauer- Büttiker Formalism, Phys. Rev. B 1992, 45, 1347.

[54] Sørensen, H. H. B.; Hansen, P. C.; Petersen, D. E.; Skelboe, S.; Stokbro, K. Efficient Wave-Function Matching Approach for Quantum Transport Calculations, Phys. Rev. B 2009, 79, 205322.

[55] Dai, X; Gao, Y. Jiang, W.; Lei, Y.; Wang, Z. U@$C_{28}$: the Electronic Structure Induced by the 32-Electron Principle. *Phys. Chem. Chem. Phys.* **2015**, *17*, 23308-23311.

[56] Xie, W.; Zhu Y.; Wang, J.; Cheng A.; Wang, Z. Magnetic coupling induced self-assembly at atomic level. *Chin. Phys. Lett.* **2019**, *36*, 116401.

[57] Liu J H, Luo K, Huang K, et al. Tunable conductance and spin filtering in twisted bilayer copper phthalocyanine molecular devices. Nanoscale Adv. 2021, 3, 3497-3501.

[58] Liu X., Wang T., Niu L., et al. Magnetic proximity, magnetoresistance and spin filtering effect in a binuclear ferric phthalo-cyanine from first principles. J. Phys. D Appl. Phys. 2019, 53, 035305.

[59] Niu L., Wang H., Bai L., et al. Spin filtering in transition-metal phthalocyanine molecules from first principles. Front. Phys. 2017, 12, 1-6.